\documentclass[aps,prl,twocolumn,amssymb,superscriptaddress]{revtex4}

\usepackage{amsmath}    \usepackage{latexsym}    \usepackage{graphicx}
\usepackage{times}

\begin{document}


\title{Black hole formation through fragmentation of toroidal polytropes}


\author{Burkhard~Zink}      \affiliation{Max-Planck-Institut     f\"ur
Astrophysik, Karl-Schwarzschild-Str. 1,  85741 Garching bei M\"unchen,
Germany}

\author{Nikolaos~Stergioulas}   \affiliation{Department   of  Physics,
Aristotle University of Thessaloniki, Thessaloniki 54124, Greece}

\author{Ian~Hawke}  \affiliation{School of Mathematics,  University of
Southampton,         Southampton         SO17         1BJ,         UK}
\affiliation{Max-Planck-Institut       f\"ur       Gravitationsphysik,
Albert-Einstein-Institut, 14476 Golm, Germany}

\author{Christian~D.~Ott}    \affiliation{Max-Planck-Institut    f\"ur
Gravitationsphysik, Albert-Einstein-Institut, 14476 Golm, Germany}

\author{Erik~Schnetter}     \affiliation{Max-Planck-Institut     f\"ur
Gravitationsphysik, Albert-Einstein-Institut, 14476 Golm, Germany}

\author{Ewald~M\"uller}     \affiliation{Max-Planck-Institut     f\"ur
Astrophysik, Karl-Schwarzschild-Str. 1,  85741 Garching bei M\"unchen,
Germany}


\date{\today}


\begin{abstract}

	We  investigate   new  paths   to  black  hole   formation  by
 considering  the general relativistic  evolution of  a differentially
 rotating polytrope with toroidal  shape.  We find that this polytrope
 is unstable to nonaxisymmetric  modes, which leads to a fragmentation
 into  self-gravitating, collapsing  components.  In  the case  of one
 such  fragment,  we  apply  a  simplified  adaptive  mesh  refinement
 technique to  follow the  evolution to the  formation of  an apparent
 horizon  centered on the  fragment. This  is the  first study  of the
 one-armed instability in full general relativity.

\end{abstract}


\maketitle


	The formation of black holes from neutron stars, iron cores or
supermassive stars is expected  to be associated with a characteristic
gravitational  wave  signal  which  may  give  information  about  the
collapse  dynamics  and  the  physical environment  of  such  objects.
Therefore,  and given  that gravitational  wave detectors  are already
taking  data  or are  coming  online, it  is  of  prime importance  to
understand  the dynamical  features of  the gravitational  collapse of
hydrodynamical systems.

	The prototypical  model of stellar collapse  is an equilibrium
polytrope subject to a  radial or quasi-radial perturbation growing on
a   dynamical  timescale.   In   spherical  symmetry,   every  general
relativistic  polytrope  with  index   $N=3$  is  unstable  to  radial
oscillations  \cite{chandrasekhar:1964b} --  in turn,  there  exists a
critical  $N_\mathrm{c}<3$ for  which the  star is  marginally stable.
Without spherical symmetry, rotation  can increase this critical value
again \cite{fowler:1966a}.  The black hole formation from the collapse
of uniformly  and differentially  rotating polytropes induced  by this
instability is a well-investigated phenomenon, either with restriction
to   axisymmetry  \cite{nakamura:1981b,   stark:1985a,  shibata:2000c,
shibata:2002c,   shibata:2003c,  shibata:2003d,   sekiguchi:2004a}  or
without      \cite{shibata:2000a,      duez:2004a,      baiotti:2004a,
baiotti:2004b}.  In the gauge  choices usually employed, the dynamical
behaviour  of the  system  shows  a radial  contraction  of the  star,
accompanied by the formation of an apparent horizon at late times.

	Black   hole    formation   from   a    dynamically   unstable
\emph{nonaxisymmetric} mode,  however, has  not been modelled  so far.
Scenarios  range  from  the  development  of a  bar  mode,  subsequent
transport  of angular  momentum into  the  shell and  collapse of  the
central object,  to fragmentation and off-center production  of one or
several   black  holes.   In   Newtonian  theory,   instabilities  and
fragmentation  have received  considerable attention,  specifically in
the   context   of    binary   formation   from   protostellar   disks
(e.g.~\cite{durisen:1986a,        tohline:1990a,        bonnell:1994a,
pickett:1996a,  banerjee:2004a} and  references  therein) and  compact
object production in  stellar core collapse (e.g.~\cite{bonnell:1995a,
davies:2002a,    shibata:2004a}   and    references    therein).    In
\cite{duez:2004a}, the  authors also report  signatures of an $m  = 4$
fragmentation behaviour  in the collapse and centrifugal  bounce of an
$N=1$  polytrope, but  could  not  determine the  final  state due  to
resolution issues.

	The   cooling   evolution  of   supermassive   stars  can   be
approximately described  by the $N=3$ mass-shedding  sequence when the
angular  momentum  transport  timescales  are short  compared  to  the
cooling timescale \cite{baumgarte:1999b},  so that uniform rotation is
enforced.   This sequence  has  a turning  point  for the  onset of  a
quasi-radial instability,  and numerical experiments  confirm that the
collapse  remains  axisymmetric \cite{saijo:2002a}.   If  the star  is
differentially rotating, the cooling  sequence is less constrained and
might   end   in   a   transition   to   nonaxisymmetric   instability
\cite{bodenheimer:1973a,   new:2001a,   new:2001b}.    The   canonical
expectation that  a supermassive star produces one  central black hole
with  a low-mass  accretion disk  might  thus not  be appropriate  for
differentially rotating configurations.

        In this  letter, we  consider the production  of a  black hole
through  the fragmentation  of a  general relativistic  polytrope.  We
focus  on $N=3$  polytropes,  which are  associated with  pre-collapse
cores of  massive stars or  supermassive stars.  The soft  equation of
state also enhances  the instability of the fragments  compared to the
common  choice $N=1$  for  neutron stars.  To  represent this  process
accurately  on a  grid, we  make use  of an  adaptive  mesh refinement
technique, since a possibly  highly deformed apparent horizon needs to
be located in some region of the domain which is unknown in advance.


	The  recent investigation  of the  collapse  of differentially
rotating supermassive stars by Saijo \cite{saijo:2004a} was based on a
sequence  of relativistic  $N  = 3$  polytropes  with a  parameterized
rotation   law  of   the   commonly  used   form   $j(\Omega)  =   A^2
(\Omega_\mathrm{c}  -  \Omega)$,   where  $\Omega_\mathrm{c}$  is  the
angular velocity  at the center,  and the parameter $A$  specifies the
degree  of differential  rotation ($A  \rightarrow \infty$  is uniform
rotation).   The  sequence  selected  was constrained  by  a  constant
central  density  $\rho_\mathrm{c} =  3.38  \times  10^{-6}$ in  units
$K=G=c=1$, and the choice $A/r_\mathrm{e} = 1/3$, where $r_\mathrm{e}$
denotes the equatorial coordinate radius.

        To  examine  the  indirect  collapse  by  fragmentation  of  a
polytrope with toroidal shape, we choose a model with the same central
density as in  Saijo's \cite{saijo:2004a} models, but with  a ratio of
polar  to equatorial  coordinate  radius $r_\mathrm{p}/r_\mathrm{e}  =
0.24$.   The  ratio  of  rotational kinetic  energy  to  gravitational
binding energy is  $T/|W| = 0.227$.  While the  critical limit for the
dynamical  f-mode instability in  uniform density,  uniformly rotating
Maclaurin    spheroids    is    ${(T/|W|)}_\mathrm{dyn}   =    0.2738$
(e.g.~\cite{tassoul:1978}), recent investigations  of the stability of
soft  ($N \sim  3$)  differentially rotating  polytropes in  Newtonian
gravity     \cite{centrella:2001a,     shibata:2002a,     saijo:2003a,
shibata:2003e}  have   shown  that  the   Maclaurin  approximation  is
inappropriate  for  such  systems,  and generally  find  the  critical
${(T/|W|)}_\mathrm{dyn}$   to   be    below   the   Maclaurin   value.
Consequently, the  toroid-like star considered in this  study might be
unstable to nonaxisymmetric perturbations.   Here we present the first
investigation of this instability  in full general relativity, showing
that relativistic effects are significant for the final outcome, as we
observe that black holes can be produced.


	All   simulations  have   been  performed   in   full  general
relativity.   The   only  assumption  on  symmetry   is  a  reflection
invariance  with respect  to the  equatorial plane  of the  star.  The
gauge freedom is fixed by  the generalized 1+log slicing condition for
the lapse function \cite{bona:1995a} with $f(\alpha)=2/\alpha$, and by
the  hyperbolic-type condition  suggested in  \cite{shibata:2003d} for
the shift vector.

	The   computational  framework   is  the   \emph{Cactus}  code
(\url{www.cactuscode.org}), which also provides  a module to solve the
geometric  part of  the field  equations in  the well-known  BSSN form
\cite{nakamura:1987a,  shibata:1995a,baumgarte:1999a}.   In  addition,
the  \emph{Carpet}  driver  \cite{schnetter:2004a}  is used  for  mesh
refinement  in \emph{Cactus}.   The  hydrodynamics part  of the  field
equations  is   evolved  using  the   high-resolution  shock-capturing
PPM-Marquina    implementation    in    the    \emph{Whisky}    module
\cite{baiotti:2003a,baiotti:2004a}, and a  gamma law equation of state
($P = \rho \epsilon/N$).  We are thus using a set of well-tested tools
to evolve the general relativistic hydrodynamics equations.

	To  numerically  construct   the  axisymmetric  initial  model
described  in the  introduction, we  use the  \emph{RNS}  initial data
solver \cite{stergioulas:1995a} with a radial resolution of 601 and an
angular resolution of 301 points. With the parameters described above,
the  model  has  toroid-like  structure, with  an  off-center  density
maximum, but a  non-zero central density.  After mapping  the model to
the hierarchy  of Cartesian grids  provided by \emph{Carpet},  a small
perturbation of the form
\begin{equation}\label{eq:ini_pert}
\rho(x)  \rightarrow \rho(x)  \Big[1 +  \frac{1}{\lambda r_\mathrm{e}}
	\sum_{m=1}^{4}  \lambda_\mathrm{m}  B  r  \sin(m  \phi)  \Big]
	\nonumber
\end{equation} 
is  applied   with  $\lambda_\mathrm{m}  =   0,  1$  and   $\lambda  =
\sum_\mathrm{i}  \lambda_\mathrm{i}$.   In  addition,  the  polytropic
constant $K$ is reduced by $0.1\%$  to induce collapse if the model is
radially  unstable.   After   perturbing  the  model,  the  constraint
equations are not solved again, since the amplitude $B$ is chosen such
that the violation  of the constraints by the  initial perturbation is
about an order of magnitude smaller than that caused by the systematic
error induced by the $m=4$ symmetry of the Cartesian grid.

	For   most  simulations,   a   fixed  \emph{box-in-box}   mesh
refinement with  5 levels  is used to  accurately resolve  the central
high-density ring. The three innermost grids cover the star, while the
two outermost  ones push the  outer boundaries to  $6.4 r_\mathrm{e}$.
The  typical resolution used  was $65  \times 65  \times 33$  per grid
patch, leading  to a central resolution  of $\delta_\mathrm{x} \approx
10^{-2} r_\mathrm{e}$.   However, runs with $89 \times  89 \times 45$,
$97 \times  97 \times 52$  and $131 \times  131 \times 65$  points per
grid patch  were also performed  to test the resolution  dependence of
the solution;  for the  last setup, a  simulation with a  uniform grid
setup would need to cover the  equatorial plane of the star alone with
320 grid points to achieve the same central resolution.

	To  determine  the  amplitude   of  a  specific  mode  in  the
equatorial  plane,  we perform  a  projection  onto  Fourier modes  at
certain coordinate radii \cite{new:2000a, devilliers:2002a}. Care must
be taken in  interpreting the results as soon as  the system starts to
deviate  significantly from axisymmetry,  since the  interpretation of
the projection  curve as a circle  assumes $\partial_\mathrm{\phi}$ to
be  a Killing  vector.   In  addition to  the  Fourier extraction,  we
monitor the evolution of the rest mass and the ADM constraints.


\begin{figure}
\centering \includegraphics[angle=0,width=8.5cm]{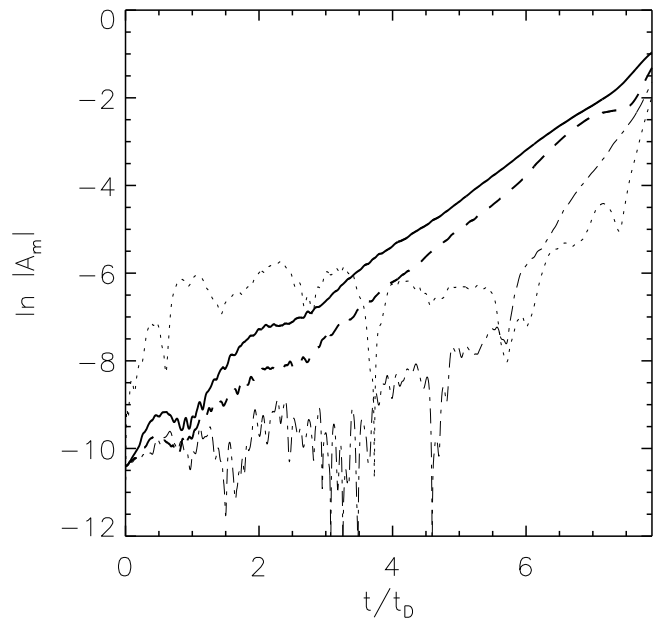}
\caption{Time evolution  of the mode  amplitudes in the  standard grid
setup N.  The amplitudes are  obtained from a Fourier decomposition of
the density profile  on the equatorial plane circle  at $\varpi = 0.25
r_\mathrm{e}$, the  initial radius of highest density.   Shown are the
$m = 1$ (\emph{thick solid line}), $m = 2$ (\emph{thick dashed line}),
$m = 3$ (\emph{thin dash-dotted  line}) and $m = 4$ (\emph{thin dotted
line})  mode amplitudes.   The  time is  normalized  to the  dynamical
timescale $t_\mathrm{D} = r_\mathrm{e} \sqrt{r_\mathrm{e}/M}$.}
\label{fig:modes_65} 
\end{figure}

	In Table~\ref{table:setups}, we have listed the parameters for
the  different simulation  runs. For  model  N, the  evolution of  the
moduli of the  equatorial Fourier components at the  initial radius of
highest  density is  shown in  Fig~\ref{fig:modes_65}.  It  is evident
that,  initially, the  $m  = 4$  component  (\emph{thin dotted  line})
induced by  our Cartesian  grid is dominant.   However, the  star is
unstable to $m = 1$ (\emph{thick solid line}) and $m = 2$ (\emph{thick
dashed line}),  and these modes  consequently grow into  the nonlinear
regime, their e-folding times being rather close.

\begin{table}
\caption{Parameters  for the simulation  runs.  Different  setups were
chosen to confirm the  results, including resolution tests, changes in
the initial perturbation, and two setups with adaptive mesh refinement
(AMR) to investigate black hole formation.}
\medskip
\begin{tabular}{c|c|c|c|c|c}
\hline & Patch &  Refinement & \multicolumn{2}{|c|}{Perturbation} & \\
Setup & resolution  & levels & Modes  & Amplitude & AMR \\  \hline N &
$65 \times 65  \times 33$ & $5$ &  $m=1-4$ & $B=10^{-3}$ & no  \\ H1 &
$89 \times 89  \times 45$ & $5$ &  $m=1-4$ & $B=10^{-3}$ & no  \\ H2 &
$97 \times 97  \times 52$ & $5$ &  $m=1-4$ & $B=10^{-3}$ & no  \\ H3 &
$131 \times 131 \times 65$ & $5$  & $m=1-4$ & $B=10^{-3}$ & no \\ I1 &
$65 \times 65  \times 33$ & $5$ &  $m=1-4$ & $B=10^{-4}$ & no  \\ I2 &
$65 \times 65 \times 33$ & $5$ &  none & $B=0$ & no \\ M1 & $65 \times
65 \times 33$ & $5-12$ & $m=1$  & $B=10^{-3}$ & yes \\ M2 & $65 \times
65 \times 33$ & $5-12$ & $m=2$ & $B=10^{-3}$ & yes \\ \hline
\end{tabular}
\label{table:setups}
\end{table}

	To test the dependence of  the results on the grid resolution,
we have evolved  the same initial data with  different grid parameters
as listed  in Table~\ref{table:setups}.  Runs  H1, H2 and H3  are high
resolution versions of N. The setups  I1 and I2 test the dependence on
the amplitude $B$  of the initial data perturbation,  with I1 using an
amplitude $10$ times  smaller than in setup N, and  I2 using $B=0$. M1
and  M2 are  variants of  N, where  only a  specific unstable  mode is
imposed.

        The  rest mass  is  conserved numerically  within $1.8\%$  for
setup  N  and  to  within   $0.2\%$  for  setup  H3.   An  approximate
measurement  of  the  e-folding  times  and mode  frequencies  can  be
obtained  within  an  error  of  $5-10\%$ related  to  ambiguities  in
defining  the  interval of  extraction.   All  setups show  consistent
results within this uncertainty.  In units of the dynamical timescale,
which    is   defined   here    as   $t_\mathrm{D}    =   r_\mathrm{e}
\sqrt{r_\mathrm{e}/M}$,   the  e-folding   times  are   $\approx  0.93
t_\mathrm{D}$ for  $m=1$, and  $\approx 0.84 t_\mathrm{D}$  for $m=2$,
respectively.   Mode frequencies  are $\approx  3.05/t_\mathrm{D}$ for
$m=1$ and $\approx 3.31/t_\mathrm{D}$ for $m=2$, respectively.

        To establish whether  a black hole is formed  by a fragment it
is necessary to cover  the fragment with significantly more resolution
than affordable by fixed mesh refinement.  Hence we have implemented a
simplified  adaptive mesh refinement  scheme to  follow the  system to
black  hole  formation: In  this  scheme,  a  tracking function,  here
provided by the location of a  density maximum, is used to construct a
locally fixed hierarchy of grids moving with the fragment.  Additional
refinement  levels  are  switched  on  during  contraction,  until  an
apparent horizon is found.

	Since the e-folding times for $m =  1$ and $m = 2$ turn out to
be close, the number and interaction behaviour of the fragments in the
non-linear  regime  depend sensitively  on  the initial  perturbation.
Thus setups M1 and M2 are  used the follow the formation and evolution
of a specific number of fragments.

\begin{figure}
\centering \includegraphics[angle=0,width=8cm]{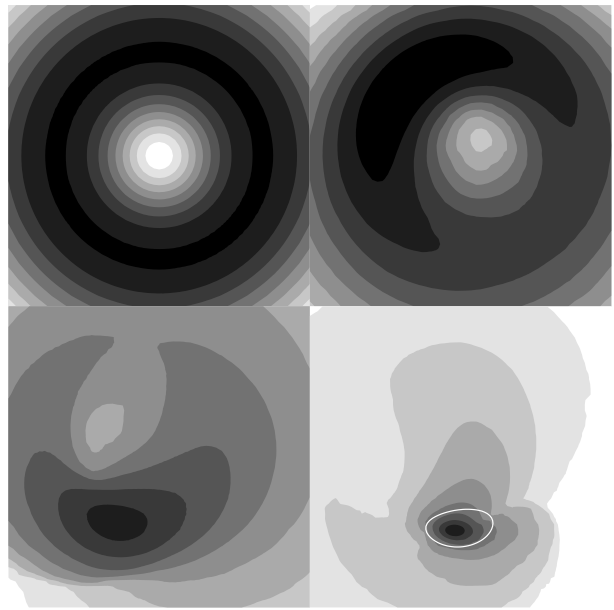}
\caption{Time evolution of the  equatorial plane density for setup M1.
Shown are isocontours of the  logarithm of the rest-mass density.  The
four  snapshots  extend  to  $0.37  r_\mathrm{e}$  and  are  taken  at
$t/t_\mathrm{D}=0$,  $6.43$, $7.14$,  and $7.45$,  respectively.  They
show the formation and collapse  of the fragment produced by the $m=1$
instability. The  last slice contains an apparent  horizon demarked by
the  thick  white  line.   Note  that  the shades  of  grey  used  for
illustration are adapted to the  current maximal density at each time,
and that darker shades denote higher densities.}
\label{fig:evol} 
\end{figure}

\begin{figure}
\centering \includegraphics[angle=0,width=8cm]{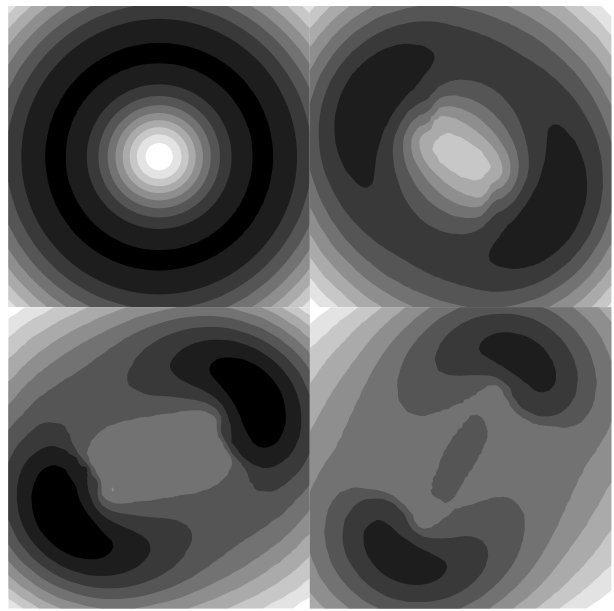}
\caption{Time evolution of the  equatorial plane density for setup M2.
The snapshot  times are  the same as  in Fig~\ref{fig:evol}.   In this
case, two fragments are  forming. Constraint violations have forced us
to terminate the simulation before apparent horizons could be located,
as explained in the text.}
\label{fig:evol_m2} 
\end{figure}

	The time  evolution of the equatorial plane  density for setup
M1  is  shown in  Fig~\ref{fig:evol}.   While  the  initial model  is
axisymmetric, it has already developed a strong $m = 1$ type deviation
from  axisymmetry  at  $t  = 6.43  t_\mathrm{D}$,  which  consequently
evolves  into  a  collapsing   off-center  fragment.   At  $t  =  7.45
t_\mathrm{D}$, we  find an apparent horizon, using  the numerical code
described in  \cite{thornburg:2004a}.  The horizon is  centered on the
collapsing fragment  at a coordinate radius  of $r_\mathrm{AH} \approx
0.16  r_\mathrm{e}$, and  has  an irreducible  mass of  $M_\mathrm{AH}
\approx  0.24 M_\mathrm{star}$.   Its  coordinate representation  is
significantly  deformed: its shape  is close  to ellipsoidal,  with an
axes  ratio of  $\sim 2:1.1:1$.   The apparent  horizon is  covered by
three refinement levels and 50 to 100 grid points along each axis.

	The    evolution   of    the    setup   M2    is   shown    in
Fig~\ref{fig:evol_m2}.   Two  orbiting  and collapsing  fragments  are
forming.  However,  even with the  adaptive mesh refinement  method we
use, constraint  violations prevent us from  continuing the simulation
to  the  formation of  apparent  horizons.   Cell-based adaptive  mesh
refinement, a  better choice of  gauge, or methods based  on numerical
analysis (e.g.  \cite{tiglio:2004a}) might be required in this case.


	To  summarize, we  have studied  fragmentation and  black hole
formation  of a  general relativistic  equilibrium polytrope  of index
$N=3$.  The  polytrope has  been shown to  be unstable  to cylindrical
perturbations of  the form $r  \sin(m \phi)$, with  $m = 1,  2$, which
consequently grow to one  or more self-gravitating fragments.  We have
applied  an adaptive  mesh  refinement method  to  resolve the  system
accurately.  In the $m = 1$ case, we have found an apparent horizon in
the spacetime, indicating that a black hole has formed.

	The dynamics of a nonaxisymmetric single-star collapse of this
type is  significantly different from  that of the  quasi-radial cases
usually investigated.  From the  often considered case of quasi-radial
collapse, to  bar formation and subsequent  collapse, to fragmentation
and fragment inspiral we have a range of possible dynamical scenarios,
which  may be connected  to discernable  observable features  in their
gravitational wave signature.  In  that sense, the evolution presented
here can be considered as an example of such processes.

	Extensive discussion  of this work as well  as further results
for the  binary case  will be presented  elsewhere. In this  case, the
fragmentation process could transform a  star into a binary black hole
merger system with a massive accretion disk around it.


       We  would like  to thank  Jonathan Thornburg  for  his apparent
horizon finder.   We used \emph{Cactus}  and the \emph{CactusEinstein}
infrastructure,  the  \emph{Whisky}  code  developed by  the  European
Network on Sources of  Gravitational Waves, and the \emph{Carpet} code
for  mesh  refinement.   In  addition,  we  are  grateful  to  Luciano
Rezzolla, Toni  Font, Anna Watts,  Nils Andersson, Motoyuki  Saijo and
Masaru Shibata for helpful comments.   This work has been supported in
part by the DFG SFB-TR7 on Gravitational Wave Astronomy.  Calculations
were performed on  the Opteron compute clusters at  the MPA, the Blade
cluster at the computing centre of the Max Planck Society in Garching,
and on the Peyote cluster of the Albert Einstein Institute in Golm.


\bibliographystyle{apsrev}


\end{document}